# Emergent Topological Hall Effect in Fe-doped Monolayer WSe$_2$

Mengqi Fang[1], Siwei Chen[1], Chunli Tang[2], Zitao Tang[1], Min-Yeong Choi[3], Jae Hyuck Jang[3], Hee-Suk Chung[3], Maya Narayanan Nair[4], Wencan Jin[2]*, and Eui-Hyeok Yang[1,5]*

[1]Department of Mechanical Engineering, Stevens Institute of Technology,

[2]Department of Physics, Auburn University

[3]Electron Microscopy Group of Materials Science, Korea Basic Science Institute

[4]Nanoscience Initiative, Advanced Science Research Cente

[5]Center for Quantum Science and Engineering, Stevens Institute of Technology

wzj0029@auburn.edu, eyang@stevens.edu Co-corresponding authors

**Abstract**
The topological Hall effect (THE) has attracted great attention since it provides an important probe of the interaction between electron and topological spin textures. THE has been considered an experimental signature of the topological spin texture of skyrmions. While THE has been widely reported in chiral magnets, oxide heterostructures, and hybrid systems such as ferromagnet/heavy metal and ferromagnet/topological insulators, the study of monolayer structures is lacking, hindering the understanding of noncollinear spin textures at the atomically thin scale. Here, we show a discernible THE via proximity coupling of Fe-doped monolayer WSe$_2$ (Fe:WSe$_2$) synthesized using chemical vapor deposition on a Pt Hall bar. Multiple characterization methods were employed to demonstrate that Fe atoms substitutionally replace W atoms, making a two-dimensional (2D) van der Waals (vdW) dilute magnetic semiconductor (DMS) at room temperature. Distinct from the intrinsic anomalous Hall effect, we found the transverse Hall resistivity of Fe:WSe$_2$ displaying two additional dip/peak features in the temperature-dependent measurements, consistent with the contribution of THE. The topological Hall effect is attributed to the magnetic skyrmions that emerge from the Dzyaloshinskii-Moriya interactions at the Fe:WSe$_2$ and Pt interface. Our work shows that a DMS synthesized from 2D vdW transition metal dichalcogenides is promising for realizing magnetic skyrmions and spintronic applications.

## 1. Introduction
The topological Hall effect (THE) has emerged as a pivotal transport property that allows probing the topological nature of electronic states beyond electrical characteristics [1–4]. The concept of THE was first introduced in the context of diluted magnetic semiconductors, distinguishing itself from the anomalous Hall effect (AHE) by its origins in spin texture irregularities rather than spin-orbit coupling [5–8]. The magnetic skyrmion, a unique spin texture with potential applications in high-speed, low-energy consumption logic and memory devices, has spurred significant interest in the study of THE [9, 10]. Their formation is typically driven by the Dzyaloshinskii-Moriya interaction (DMI), a chiral exchange interaction encouraged by broken inversion symmetry [11–14]. DMI is instrumental in crafting chiral domain walls and skyrmions, observed in various noncentrosymmetric chiral lattice magnets [15, 16].

THE has been widely reported in chiral magnets and hybrid systems, including complex oxide heterostructures [17–19] and ferromagnet/heavy metal multilayers [20]. Among them, robust magnetic ordering and substantial spin-orbit coupling (SOC) have been recognized as an effective strategy to induce DMI. In this context, two-dimensional (2D) van der Waals (vdW) materials provide a fertile ground for exploring THE, associated with topological spin textures. Recently, Néel-type skyrmions were shown to exist in mechanically exfoliated 1T′-$WTe_2$/$Fe_3GeTe_2$ heterostructures [21]. A possible presence of antisymmetric peaks in $Cr_{1.2}Te_2$ sheets was reported at temperatures up to 320 K [22]. A giant THE signal of 1.39 μΩ·cm was observed in the vdW heterostructures of $CrTe_2$/$Bi_2Te_3$ [23]. Despite these developments, the exploration of THE in atomically thin monolayers is as yet underexplored.

Recently, substitutional doping of monolayer transition metal dichalcogenides (TMDs) synthesized via chemical vapor deposition (CVD) has emerged to introduce magnetic impurities to replace transition metal atoms within the lattice for covalent integration [24–26]. For instance, Fe-doped $MoS_2$ [27–30] and $MoSe_2$ [31] monolayers, V-doped 2H-$MoTe_2$ [32], V-doped $WS_2$ [33] and $WSe_2$ monolayers [34] have been shown to persist ferromagnetism at room temperature. The resultant 2D vdW dilute magnetic semiconductors (DMS) provide a new route for implementing long-range magnetic order at room temperature. Meanwhile, monolayer TMDs exhibit strong SOC [35, 36] owing to the heavy transition metals they contain. Therefore, 2D vdW DMS synthesized from monolayer TMDs are promising candidates for exploring DMI-induced topological spin textures. In particular, 2D vdW DMS synthesized via Fe-doping of $WSe_2$ monolayers would favor the existence of skyrmion since the closest Se atoms to the Fe atoms would have opposite spin coupling, while the nearby W atoms would have the same spin coupling [37].

Here, we show a discernible THE via proximity coupling of Fe-doped monolayer $WSe_2$ (Fe:$WSe_2$) synthesized using chemical vapor deposition on a Pt Hall bar and temperature-dependent Hall resistivity measurements. We first demonstrate room-temperature ferromagnetism in Fe:$WSe_2$ monolayers synthesized via CVD. The photoluminescence (PL) spectra and first principles calculation reveal the spin-polarized electronic structure due to Fe-doping. The Fe doping concentration is determined using X-ray photoelectron spectroscopy (XPS) and scanning transmission electron microscopy (STEM), confirming that Fe:$WSe_2$ monolayers are a 2D DMS. Remarkably, two additional dip/peak features emerge in the Hall resistivity experiment, which is distinct from the anomalous Hall effect (AHE). Such features are consistent with the signal from THE, indicating the presence of skyrmions. Our work provides a scalable platform to explore and manipulate spin textures in atomically thin materials and interfaces, paving the way for developing nanoscale spintronic devices.

## 2. Results

Fe:WSe$_2$ and undoped WSe$_2$ monolayers were synthesized via CVD. The schematic of the CVD system for Fe:WSe$_2$ and WSe$_2$ growth and the process parameters are described in Figures 1a and S1. Figure 1b shows the schematic of atomic structure of monolayer Fe:MoS$_2$. Scanning electron microscopy (SEM) image shows that Fe:WSe$_2$ monolayers are crystalized in an equilateral triangular shape with an average domain size of ~15 μm (Figure S2a). The atomic force microscopy (AFM) image in Figure S2b provides a topographical representation of the Fe:WSe$_2$, with a distinct lattice height measured at approximately 0.86 nm, probing it as a monolayer thickness [38]. Figures 1c and S2c,d show high-resolution transmission electron microscopy (HRTEM) depiction of monolayer Fe:WSe$_2$, depicting the hexagonal lattice architecture and crystalline integrity of the (100) plane, with an interplanar spacing of ~0.27 nm. The Fast Fourier Transform (FFT) pattern, as observed in Figure 1d, corroborates the crystalline periodicity and the spatial orientation of the monolayer structure. Figure 1e depicts a scanning transmission electron microscopy (HAADF-STEM) image, showing darker spots marked by a red arrow (Fe atom) in the bright white (W atoms) atomic lattice, indicating a Fe atom substituted a W atom. The corresponding cross-sectional intensity profile in Figure 1f reveals the contrast ratio between Fe and W atoms, corresponding to the atomic mass difference. This result fits the simulation intensity profile shown in Figure 1f.

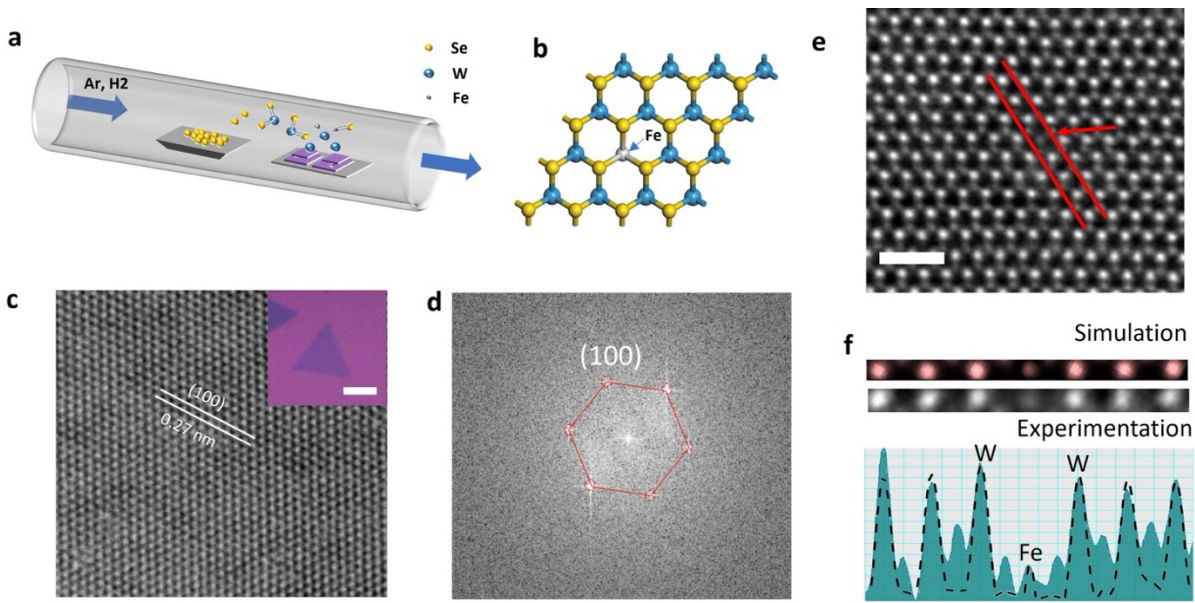

**Figure 1. Structural characterizations of Fe:WSe$_2$ and undoped WSe$_2$ monolayer crystals. a)** Schematic of synthesis of Fe-doped WSe$_2$ using chemical vapor deposition (CVD). **b)** Schematic of atomic structure of Fe:MoS$_2$, where tungsten atoms appear in blue, selenium atoms in yellow, and iron atom in gray. **c)** HRTEM image of as-grown Fe:WSe$_2$. Insert: optical image of Fe:WSe$_2$. **d)** Fast Fourier transform (FFT) pattern of the area shown in **(c)**. **e)** HAADF-STEM atomic image of Fe:WSe$_2$ and **f)** corresponding intensity profile. Scare bar in e) is 1 nm.

To identify the location of Fe atoms within the monolayer $WSe_2$, high-angle annular dark-field XPS measurement examined the disparities in chemical structure and energy levels between Fe-doped and undoped $WSe_2$ as depicted in Figure S3a, revealing distinct Fe 2p3 peaks at 708.465 eV, aligning with previously documented findings [39, 40]. The quantification of atomic concentration unveiled a range of 0.05% to 0.3%. Our synthesis method exhibited limited tunability in doping concentration, primarily attributed to the nature of our growth method, where transition metal atoms are supplied from a thin metal film to promote the controlled sublimation of $MoO_3$ [27, 41]. Figures S3b and S3c present the binding energy profiles for W 4f and Se 3d in Fe:$WSe_2$ and $WSe_2$, respectively. For Fe:$WSe_2$, the W 4f binding energy spectrum features two conspicuous peaks at 32.716 eV for W $4f_{7/2}$ and 34.839 eV for W $4f_{5/2}$. These values indicate an upshift of approximately ~0.202 eV and ~0.125 eV, respectively, compared to the binding energies observed in undoped $WSe_2$ (32.514 eV and 34.714 eV, respectively). Moreover, two discernible W 6+ peaks, positioned at approximately ~36.122 eV and 38.148 eV, are noted, attributed to $WO_3$-related features. Intriguingly, the intensity of the W 6+ peaks diminishes after introducing Fe into the crystal lattice, hinting at a concomitant reduction in the selenium vacancy concentration resulting from the doping process [27]. Likewise, the binding energy of Se 3d core levels within the Fe-doped sample showcases a consistent upshift of approximately ~0.176 eV for Se $3d_{3/2}$ and ~0.195 eV for Se $3d_{5/2}$ when contrasted with the undoped counterpart, which is attributed to lattice distortions induced by the introduction of Fe atoms into the crystal lattice of $WSe_2$. These results show that monolayer $WSe_2$ with occupation sites of Fe atoms is successfully obtained, where Fe atoms substitute for W in the monolayer $WSe_2$ host. The Raman and photoluminescence (PL) spectra of monolayers Fe:$WSe_2$ and $WSe_2$ are shown in Figure S4, and their analysis is described in detail in Note 1.

To study the impact of Fe dopants on the magnetic properties of $WSe_2$, we performed temperature-dependent vibrating sample magnetometer (VSM) measurements. Figure 2a shows the magnetic hysteresis loops of Fe:$WSe_2$ and $WSe_2$ monolayers measured at 5K, respectively. The Fe:$WSe_2$ displays a pronounced magnetic response compared to the undoped sample, evident from the substantial magnetic moment in Fe:$WSe_2$ over the entire magnetic field range. A clear hysteresis loop in Fe:$WSe_2$ suggests a ferromagnetic behavior, whereas the undoped sample exhibits negligible magnetic moment, indicating its diamagnetic nature. The temperature-dependent *M-H* loops for Fe:$WSe_2$ and $WSe_2$ are shown in Figures S5 and S6, respectively. Figure S7 shows that $H_c$ changes with temperature for both Fe:$WSe_2$ and $WSe_2$ monolayers, exhibiting a clear difference in the coercivity trends between them. The Fe:$WSe_2$ shows variable coercivity with temperature, while the undoped $WSe_2$ coercivity remains relatively low. They both display similar $H_c$ values at a low temperature, potentially indicating a characteristic magnetic behavior intrinsic to $WSe_2$ irrespective of doping. However, they manifest divergent coercivity trends as the temperature rises, which indicates their differential impact of thermal energy on the magnetic properties, attributed to the exchange interactions with rising thermal energy [42, 43] The saturation magnetization ($M_s$) against temperature for both Fe:$WSe_2$ and $WSe_2$ monolayers is shown in Figure 2b. Observably, the monolayer Fe:$WSe_2$ boasts a higher $M_s$ across the temperature range, reaching its zenith around 120 K, after which it experiences a decline, which is typical of many magnetic materials due to increased thermal agitation affecting the alignment of magnetic moments. The undoped $WSe_2$, on the other hand, showcases a relatively subdued saturation magnetization. The greater $M_s$ in Fe:WSe$_2$ reiterates the efficacy of Fe in bolstering the magnetic character of WSe$_2$.

We performed comprehensive density functional theory (DFT) calculations on the electronic structure of monolayers Fe:WSe$_2$ and WSe$_2$ to elucidate the magnetic properties of the Fe-doped system . Figures 2c and 2e depict the schematic representations of monolayers Fe:WSe$_2$ and WSe$_2$, respectively, highlighting the atomic arrangement and potential sites for Fe incorporation. Figures 2d and 2f present the corresponding density of states (DOS) for both spin-up (majority) and spin-down (minority) electron states from -4 eV to 4 eV. In the DOS plot for Fe:WSe$_2$ (Figures 2d), a pronounced spin polarization is evident, as indicated by the significant disparity between the spin-up and spin-down states across the low Fermi energy range. This spin polarization is a direct consequence of the Fe doping, which introduces localized magnetic moments within the WSe$_2$ lattice. Near the Fermi level (-0.25~0 eV), the DOS for spin-up states is significantly higher than that for spin-down states, suggesting that the majority of charge carriers contributing to the magnetic moment are spin-polarized. This imbalance between spin-up and spin-down states is a hallmark of ferromagnetic materials, where the alignment of electron spins in a specific direction results in a net magnetic moment. The exchange splitting observed between these states is attributed to the interaction between the *d*-electrons of the Fe atoms and the *p*-electrons of the W atoms in the WSe$_2$ lattice. This interaction induces local magnetic moments at the Fe sites and facilitates a long-range ferromagnetic order within the monolayer.

Moreover, the magnitude of spin polarization near the Fermi level in Fe:WSe$_2$ indicates a strong coupling between the electronic and magnetic properties, which could be pivotal in determining the suitability of materials for spintronic applications. The presence of Fe in the WSe$_2$ lattice disrupts the otherwise symmetric electronic structure of the host material, leading to the observed asymmetry in the DOS. This asymmetry shows the significant role of Fe doping in modulating the electronic and magnetic characteristics of WSe$_2$. In contrast, the DOS for the undoped WSe$_2$ (Figure 2f) shows a balanced distribution of spin-up and spin-down states, indicative of a non-magnetic material. This balance is typical for WSe$_2$, which, in its pristine form, does not exhibit any intrinsic magnetic ordering.

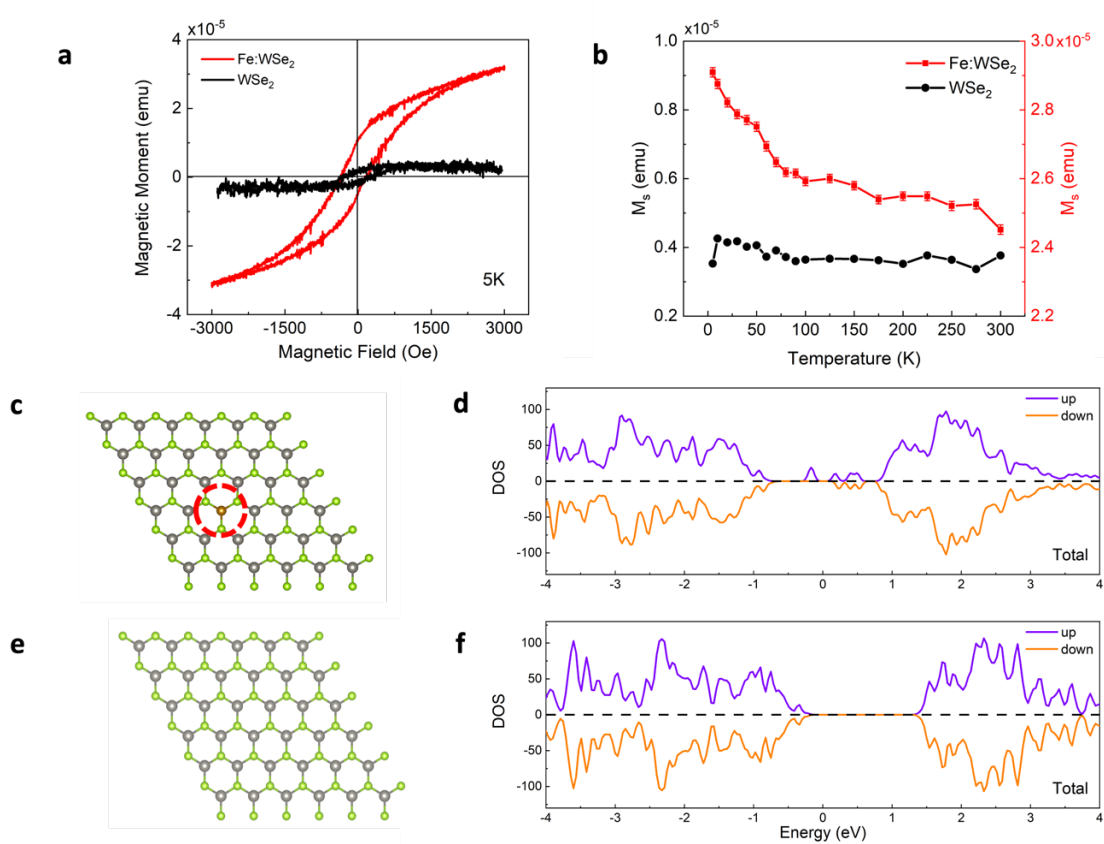

**Figure 2. Magnetic properties of Fe:WSe$_2$ and WSe$_2$ monolayer. a)** Magnetic hysteresis loops for WSe$_2$ and Fe:WSe$_2$ at 5 K, showing pronounced ferromagnetic behavior in the Fe-doped sample as evidenced by the loop width. **b)** Temperature-dependent saturation magnetization ($M_s$) curves for WSe$_2$ and Fe:WSe$_2$, illustrating the retention of magnetic order in the doped material up to 300 K. **c)** Schematic representation of a WSe$_2$ monolayer with a Fe atom substitutionally doped at a W site, highlighted by the red dashed circle. **d)** Density of states (DOS) for Fe:WSe$_2$, showing spin-polarized electronic states with a clear distinction between spin-up and spin-down contributions near 0 eV. **e)** Schematic representation of an undoped WSe$_2$ monolayer lattice structure. **f)** DOS for undoped WSe$_2$, displaying a nearly overlapping profile for spin-up and spin-down states near 0 eV.

We furthermore characterized the proximate effect of a Fe:WSe$_2$ monolayer on the platinum (Pt) Hall bar. Figure 3a shows the schematic depicting the monolayer Fe:WSe$_2$/Pt Hall bar heterostructure device for Hall resistivity measurements. The Hall voltage was measured across the perpendicular direction, which is markedly influenced by contributions from the ordinary Hall effect (OHE), AHE, and THE (Figure 3b). After subtracting the OHE, we decompose the transverse Hall resistance ($R_{xy}$) as $R_{xy} = R_{AHE} + R_{THE}$, where $R_{AHE}$ and $R_{THE}$ denote anomalous Hall effect and topological Hall effect, respectively [7, 44]. Figure 3c shows the AHE contribution subtracted from Figure 3b data using $M_0 \tanh\left(\frac{H}{a_0} - H_0\right)$ [17, 23]. After the $R_{AHE}$ subtraction, $R_{xy}$ measured at 5 K exhibits a positive maximum of 7.5 mΩ for a field of about 45 Oe when the field is swept from +1 T to -1 T, while antisymmetric peak occurs in the opposite field sweeping direction. The observation of antisymmetric spikes strongly suggests the manifestation of THE, resembling the presence of skyrmion phases similar to those observed in other heavy metal (HM)/ferromagnet (FM) heterostructures, including Pt/Tm$_3$Fe$_5$O$_{12}$ [20, 45]. The skyrmions in HM/FM structures are known to be generated by the interfacial Dzyaloshinskii-Moriya interaction (DMI). As electrons traverse through the material, they interact with the skyrmion texture by aligning their spins with the emergent local magnetic configuration (Figure 3d). This interaction gives rise to the transverse Hall voltage, which manifests as the characteristic humps [10, 46, 47]. The presence of skyrmions, characterized by their intricate spin textures, suggests the existence of robust SOC at the interface between the 2D Fe:WSe$_2$ and the Pt [48]. Notably, $R_{xy}$ measured at 5 K in the Hall bar device of WSe$_2$/Pt does not exhibit such spikes (Figure S9 above), which confirms that Fe doping into monolayer WSe$_2$ contributes to the emergence of skyrmions within this system, fostering a noncoplanar spin arrangement [49, 50]. Note that neither AHE nor THE was observed in the monolayer Fe:WSe$_2$/Au heterostructure (Figure S9 bottom). This is attributed to the weak SOC between Fe:WSe$_2$ and Au, suppressing interfacial DMI-induced THE. Although similar spikes have been reported as a result of surface and bulk ferromagnetic components in multi-layered magnetic structures, similar to the (V,Bi,Sb)$_2$Te$_3$/ (Bi,Sb)$_2$Te$_3$ structure [51], the Fe:WSe2 used here is a monolayer, where the contribution from the bulk ferromagnetic component is minimal, making such a case unlikely.

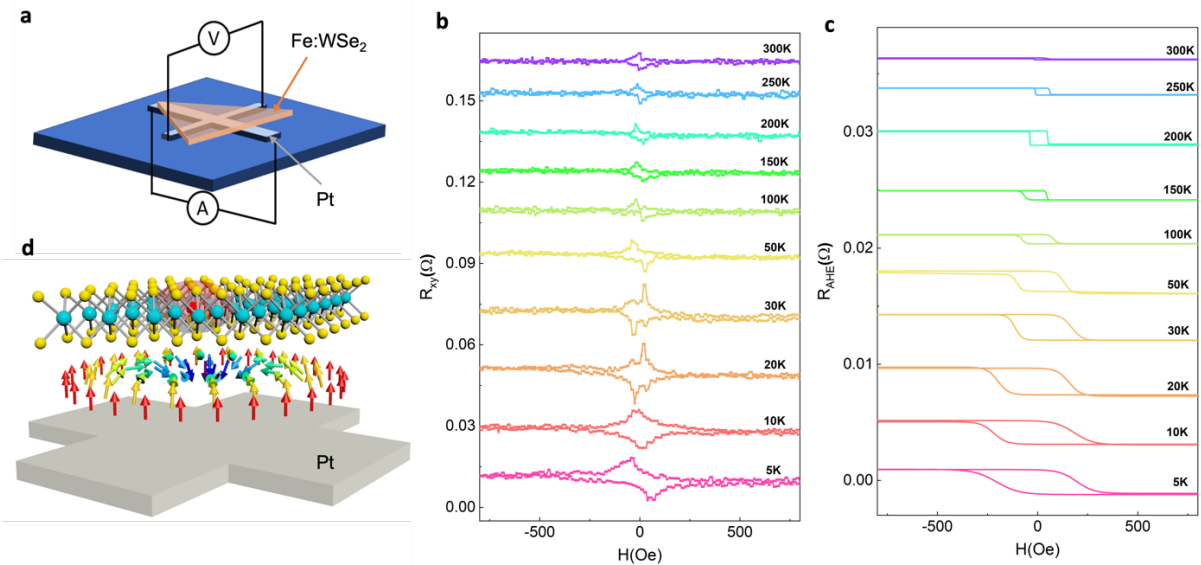

**Figure 3. Topological Hall effect (THE) measurement of monolayer Fe:WSe$_2$.** a) Illustration of the Hall effect measurement in monolayer Fe:WSe$_2$ crystal on a Pt bar. b) Temperature-dependent Hall resistivity ($R_{xy}$) of monolayer Fe:WSe$_2$, demonstrating discernible topological Hall effect. c) Fitted Anomalous Hall resistivity at different temperatures. d) Schematic of skyrmion texture in Fe:WSe$_2$/Pt layer with a vdW gap.

Figure 4a shows the decomposed Hall resistance with contributions from both AHE and THE at 5 K. The AHE contribution, highlighted in dark blue, is weak compared to that of THE, suggesting that the proximity-induced magnetization by Fe:WSe$_2$ within the Pt is relatively weak [52–54]. The AHE traditionally depends on a scattering characteristic of highly conductive systems [53, 55]; however, such conditions are challenging to achieve within our experimental setup due to the potential for heat generation within the Pt layer since excessive heat could damage the monolayer Fe:WSe$_2$.

Figure 4b shows the Hall measurement at 20 K, where two double humps appear (marked as fit 1 and fit 2). Here, the 'cross-over' peaks marked as fit are attributed to the formation of a noncoplanar spin texture during a weak magnetic field, which results from the interplay between magnetocrystalline anisotropy and ferromagnetic interactions [56]. The ratio of uniaxial anisotropy to shape anisotropy is described by $Q = \frac{2K}{\mu_0 M_s^2}$, where $K$ is the magnetic anisotropy constant and $M_s$ is the saturation magnetization [57]. While $M_s$ is strong at 5 K, resulting in large shape anisotropy, it decreases at 20 K and 30 K as thermal energy disrupts the alignment of magnetic moments and interlayer exchange coupling weakens [58, 59]. This decrease in $M_s$ facilitates the formation of a donut-like magnetic spin state (donut state) around a central hole [57, 60]. However, as the temperature further rises, the uniaxial anisotropy weakens [61], causing this donut state to revert to a single skyrmion state, which remains stable up to 300 K.

Figure 4c depicts the temperature dependence of the topological Hall resistance ($R_{THE}$) and the anomalous Hall resistance ($R_{AHE}$), respectively. At 5 K, $R_{THE}$ (denoted by the blue squares) is at its highest (7.5 mΩ), indicating strong THE due to robust skyrmion formation. As the temperature rises, $R_{THE}$ experiences a sharp drop around 20 - 30 K, followed by a more gradual decline up to 300 K. This trend suggests that the skyrmion-induced THE diminishes with increasing thermal energy and persists even at room temperature. Moreover, $R_{AHE}$ (shown by the red triangles) also decreases with increasing temperature. Starting from 1.2 mΩ at 5 K, $R_{AHE}$ gradually decreases as the temperature rises, indicating a slight weakening of the AHE. The relatively lower magnitude and less steep decline of $R_{AHE}$ compared to $R_{THE}$ is attributed to the proximity-induced magnetization in the Pt layer, which plays a less important role in the overall Hall response than the topological contributions. The temperature profile of the topological Hall resistivity peak, $H_T$, commences at its zenith value of 44 Oe at 5 K (Figure 4d). As the temperature increments, $H_T$ experiences a gradual reduction, showcasing a zone of fluctuation between 100 K and 200 K before facing a decline up to 300 K. This trend resembles the behavior of coercive field ($H_c$) captured by VSM in Figure 2b. The stable room temperature magnetic phase holds promise for future applications in logic and memory devices. [48, 62]. Figures 4e and 4f show the temperature-magnetic field phase diagram integrating the robust THE response over the entire temperature range. Two signals appearing at opposite fields at temperatures from 20 K to 30 K indicate the presence of a complex skyrmion phase. The magnitude of the topological Hall resistance, $R_{THE}$, persists up to 300 K, indicating that the robust magnetic skyrmion is maintained over a wide temperature range and that the topological features at the Fe:MoS$_2$/Pt interface are stable.

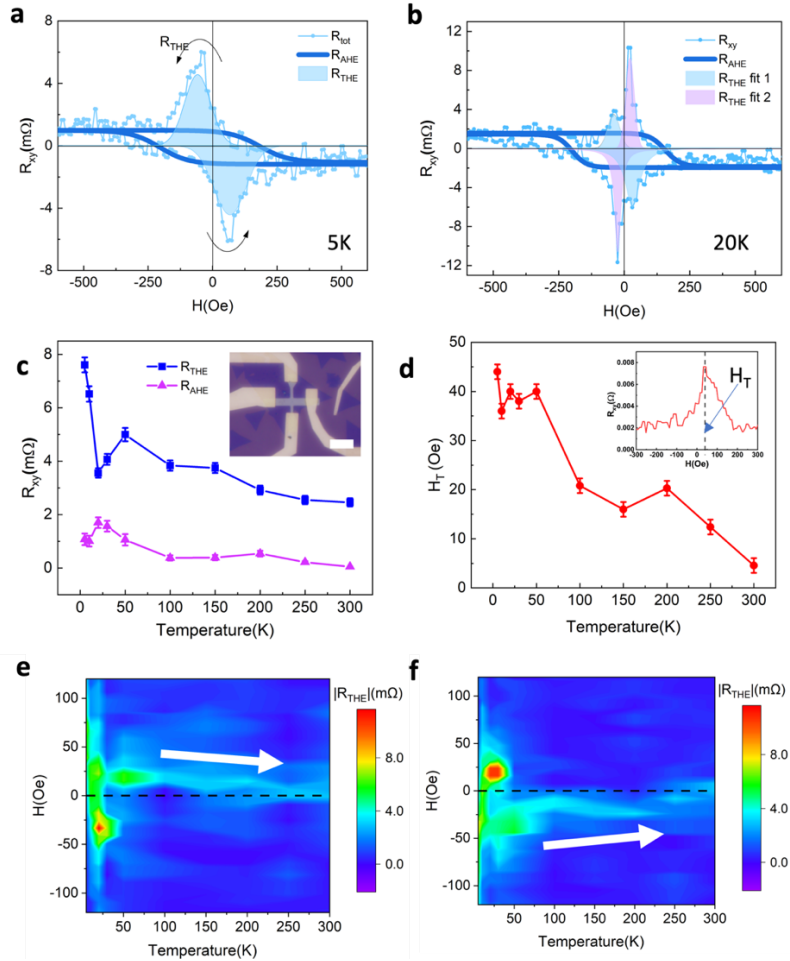

**Figure 4. Temperature-dependent of topological Hall effect.** a) and b) Decompositions of $R_{xy}$ peaks by subtracting the ordinary Hall contributions. Contributions from AHE (dark blue lines) and THE (blue and purple area) are marked. c) Magnitudes of the topological Hall resistance ($R_{THE}$) and the anomalous Hall resistance ($R_{AHE}$), respectively, across a range of temperatures. Inset: Optical image of the measured Fe:WSe$_2$/Pt heterostructure, with a scale bar representing 5 μm. d) Field strengths at which the topological Hall resistivity reaches its maximum ($H_T$) as a function of temperature. Inset: Peak position correlating to the maximum peak. e) and f) Temperature-magnetic field dependence of THE resistance response with external magnetic fields changed from -1 T to 1 T and 1 T to -1 T.

## 3. Conclusion

We have synthesized WSe$_2$ monolayers with Fe atoms substitutionally replaced W atoms via CVD and verified their temperature-dependent ferromagnetic properties, which persist up to room temperature. The XPS and STEM analyses confirmed the doping concentration of 0.3%, while its unique PL shift and quenching indicate changes in the electronic structure due to Fe doping. Through VSM and Hall effect measurements, magnetic characterizations affirmed the temperature-dependent magnetic properties and highlighted the existence of intriguing magnetic textures, possibly due to skyrmionic structures. THE in Fe:$WSe_2$ monolayers highlight the potential for manipulating magnetic textures. The robustness of THE persistent up to room temperatures suggests the existence of stable induced skyrmion-like features. Complex spin textures from Fe doping, which are crucial for THE, are less affected by the thinness or interface irregularities that can weaken AHE. This research enhances our grasp of both ferromagnetism and the interplay between topological and anomalous Hall effects in 2D vdW DMS systems, paving the way for creating scalable and integrated heterostructures into spintronic and non-volatile memory applications.

## 4. Methods

*Synthesis of normal and Fe-doped $WSe_2$*

A 5 nm-thick $WO_3$ (99.99%, Kurt J. Lesker) was deposited onto a $SiO_2$/Si wafer via a joule heating, serving as the transition metal source. $Fe_3O_4$ particles (Alpha Chemicals) and sulfur or selenium powders (99.5% Alfa Aesar) were used as Fe and Se sources, respectively. Before the growth, the $SiO_2$/Si substrate was immersed in a 10% KOH solution for 3 minutes to increase surface energy, followed by a DI water treatment. Then, the $Fe_3O_4$ particles were placed on the $SiO_2$/Si substrate and brought into contact with the $WO_3$-deposited substrate. Subsequently, the Se was provided by vaporizing Se powders placed on crucibles outside the central furnace area as the temperature increased to 690 °C. During this process, Ar gas was flown into the furnace at 50 °C, while $H_2$ gas was introduced into the furnace at 690 °C. The furnace was then heated to a maximum temperature of 850 °C and held at this temperature for 15 minutes. Then, the hydrogen gas supply was halted while maintaining Ar gas flow until the furnace cooled down to room temperature. After the process, monolayer Fe-doped TMDs were formed. The growth process of undoped $WSe_2$ monolayers closely resembled that of Fe:$WSe_2$ monolayers, except that $Fe_3O_4$ was not incorporated during growth.

*Sample preparation for characterization*

To transfer Fe:$WSe_2$ monolayer on a lacey carbon TEM grid, Fe:$WSe_2$ and undoped $WSe_2$ samples on a $SiO_2$/Si substrate was coated with PMMA 950 A4 using a dropper and left to air-dry at room temperature for 1 hour. The substrate was then submerged in a 30% KOH solution, causing the Si substrate to detach from PMMA after 30 minutes, with the PMMA and $WSe_2$ layers remaining afloat. The PMMA was cleaned in DI water and blow-dried with air. The PMMA-coated $WSe_2$ layer was transferred onto the lacey carbon TEM grid with gentle pressure applied using a glass slide. Then, the PMMA was removed by warm acetone at 45 °C for 30 min, then annealed in a low-pressure vacuum chamber for 3 hours ($WSe_2$ and Fe:$WSe_2$) at 250 °C to fully remove the polymer residuals.

*Sample Characterization*

**AFM:** AFM and MFM analyses were conducted using a Bruker BioScope Resolve atomic force microscope. Measurements took place under atmospheric conditions. For MFM observations, Co-Cr coated MESP-V2 tips (Bruker) in lift mode were used, and the tips' magnetization was achieved using a permanent magnet.

**SEM:** SEM observations were carried out using Zeiss Auriga Small Dual-Beam FIB-SEM. This microscope has a GEMINI field-emission electron column with a resolution of 1.0 nm @ 15 kV and 1.9 nm @ 1 kV. The SEM images were acquired at 3 kV with the in-lens EsB (energy selective backscattering) detector.

**STEM:** The atomic resolution Scanning Transmission Electron Microscopy (STEM) image was observed using a monochromated Cs-corrected Transmission Electron Microscope (TEM) with a monochromated ARM200F(JEOL).

**Raman and PL:** Raman and photoluminescence (PL) spectroscopy measurements were conducted using a Horiba XPLORA spectrometer fitted with a 532 nm laser, operating at atmospheric conditions.

**XPS:** The X-ray photoelectron spectroscopy (XPS) data were gathered using a system equipped with a monochromatic aluminum Kα X-ray source (1486.6 eV) and a hemispherical energy analyzer. The carbon 1s peak at 284.8 eV was a reference for binding energy calibration. XPS peak deconvolution was carried out using spectral analysis software that utilizes a Voigt profile model along with Shirley-type background subtraction to fit the data.

*Magnetic measurements*

The magnetization and transport measurements were carried out using a commercial physical property measurement system (PPMS, Quantum Design DynaCool, 1.8K-400K, 9T). The magnetization was measured using a VSM. The Hall resistivity was acquired in a 4-terminal Hall bar device option at various temperatures and fields. The DC current was supplied from a Keithley 6221 DC and AC current source, and the signal was read by a Keithley 2182A nanovoltmeter.

*Device fabrication*

The substrate was coated with PMMA and then prebaked at 180 °C for 90 seconds. Following prebaking, e-beam exposure developed the photoresist using a 3:1 solution of 4-methyl-2-pentanone (MIBK) to 2-propanol (IPA). After the development step, a Pt/Cr film, approximately 8/5 nm thick, was deposited onto the patterned substrate via Joule heating. The patterning was completed via a lift-off process, which entailed rinsing with acetone and IPA to define the Pt/Cr electrodes. This lithography process was repeated to align an additional electrode layer with the pre-existing Pt pattern, onto which an Au/Ar layer, roughly 50/10 nm in thickness, was added. Following the second lift-off step, Fe:WSe$_2$ monolayers were placed atop the Pt electrodes via a wet transfer technique. The method concluded with attaching Al wires to the gold pads and finalizing the electrode assembly for subsequent electrical measurements.

*DFT calculations*

The DFT calculations were carried out using the Vienna ab initio simulation (VASP5.4.4) code based on the plane-wave basis sets with the projector augmented-wave method [63, 64]. The generalized gradient approximation (GGA) was applied for the exchange-correlation potential following the Perdew-Burke-Ernzerhof scheme [65, 66]. An energy cutoff of 400 eV and a Γ-centered Monkhorst-Pack grid of 6 × 6 × 1 were used for Brillouin-zone sampling [67]. The criteria for energy convergence were set to $10^{-5}$ eV.

**Acknowledgments**

This work was supported in part by the National Science Foundation award (ECCS-1104870) and the Defense University Research Instrumentation Program (FA9550-11-1-0272). This research used microscopy resources, including AFM, SEM, and HRTEM, partially funded by the NSF via Grant NSF-DMR-0922522, within the Laboratory for Multiscale Imaging (LMSI) at Stevens Institute of Technology. This work was also partially carried out at the Micro Device Laboratory (MDL) at Stevens Institute of Technology, funded with support from W15QKN-05-D-0011. C.T. and W. J. acknowledge support from the National Science Foundation Grant No. DMR-2129879 and DMR-2339615.